\documentclass[]{aa}
\usepackage{graphicx}
\usepackage{natbib}
\bibpunct{(}{)}{;}{a}{}{,}
\include{epsf}
\include{rotate}

\begin{document}

\title{ 
Moment equations for chemical reactions 
on interstellar dust grains
}
\author{
Azi Lipshtat 
\and 
Ofer Biham}  


\institute{
Racah Institute of Physics, 
The Hebrew University, 
Jerusalem 91904, Israel 
}

\date{Received ; Accepted }

\abstract{
While most chemical reactions in the interstellar medium take place in the gas phase,
those occurring on the surfaces of dust grains play an essential role.
Such surface reactions include the catalytic production of molecular hydrogen 
as well as more complex reaction networks producing 
ice mantles and various organic molecules.
Chemical models based on rate equations 
including both gas phase and grain surface reactions have been used
in order 
to simulate the formation of chemical complexity in interstellar clouds. 
For reactions in the gas phase  
and on large grains,
rate equations, 
which are highly efficient to simulate,
are an ideal tool.
However, for
small grains 
under low flux,
the typical 
number of atoms or molecules of certain reactive species
on a grain may go down to order one or less.
In this case 
the discrete nature of the populations of reactive species as well as 
the fluctuations become dominant, thus
the mean-field approximation 
on which the rate equations are based does not apply. 
Recently, a master equation approach, that provides a good
description of chemical reactions on interstellar dust grains,
was proposed. 
Here we present a related approach based on {\it moment equations}
that can be obtained from the master equation. 
These equations describe the time evolution 
of the moments of the distribution of the population
of the various chemical species
on the grain. 
An advantage of this approach is the fact that
the production rates of molecular species are expressed directly in terms of
these moments.
Here we use the moment equations to calculate the rate of
molecular hydrogen formation on small grains.
It is shown that the moment equation approach is efficient
in this case in which only a single reactive specie is involved. 
The set of equations for the case of two species is presented
and the difficulties in implementing this approach for complex
reaction networks involving multiple species are discussed.
}

\titlerunning{Moment equations for reactions on dust grains}
\authorrunning{Lipshtat \& Biham}

\maketitle

\newpage

\section{Introduction}
        \label{intro.}

The chemistry of interstellar clouds consists of reactions taking place in
the gas phase as well as on the surfaces of dust grains
\citep{Hartquist1995}.
Reactions that take place on dust grain surfaces include the formation
of molecular hydrogen as well as reaction networks producing ice
mantles and various organic molecules.
The formation of H$_2$ on dust grain surfaces is a process of fundamental
importance due to the fact that H$_2$ cannot form in the gas phase efficiently
enough to account for its observed abundance
\citep{Gould1963,Hollenbach1970,Hollenbach1971a,Hollenbach1971b}.
Molecular hydrogen plays a crucial role 
in gravitational collapse and star formation by absorbing energetic photons
and radiating the energy in the infrared, to which the cloud is more transparent,
thus cooling it and enabling further collapse.
Furthermore, H$_2$ molecules are a necessary component for the initiation of chemical reaction
networks that give rise to the chemical complexity observed in interstellar  
clouds.
Therefore, the process of H$_2$ formation has attracted the attention of both
theorists
\citep{Williams1968,Smoluchowski1981,Smoluchowski1983,Aronowitz1985,Duley1986,Pirronello1988,Sandford1993,Takahashi1999} 
and 
experimentalists
\citep{Brackmann1961,Schutte1976,Pirronello1997a,Pirronello1997b,Pirronello1999,Manico2001}.

The computational modeling of 
chemical reaction networks on dust grains in the interstellar medium
is typically done using rate
equation models
\citep{Pickles1977,d'Hendecourt1985,Brown1990,Brown1990b,Hasegawa1992,Hasegawa1993a,Hasegawa1993b,Caselli1993,Willacy1993,Shalabiea1994}. 
These models consist of coupled ordinary differential equations that
provide the time derivatives of the densities of the species involved.
Integration of these equations provides the time evolution of the densities.
As long as the grains are not too small and the flux is not too low,
the mean-field approximation applies. 
In this case, rate equations are an ideal tool for the simulation of 
surface
reactions, due to their high computational efficiency.
However, in the limit of very small grains under very low flux, 
rate equations are not always valid.
This is because they
take into account only average densities and ignore the fluctuations as
well as the discrete nature of the populations of the atomic and molecular species
\citep{Tielens1995,Charnley1997,Caselli1998,Shalabiea1998,Stantcheva2001}. 
These features become significant in the limit of small grains
and low incoming flux,
typically encountered in 
diffuse interstellar clouds where hydrogen recombination 
as well as various other grain surface reactions
take place.
For example, 
as the number of H atoms on a grain fluctuates in the range of 0, 1 or 2,
the H$_2$ formation rate cannot be obtained from the average 
number alone.
This can be easily understood, since the recombination process requires at least two
H atoms simultaneously on the surface.

Recently, a master equation approach was proposed, that 
is suitable for the simulation of chemical reactions on
microscopic grains
\citep{Biham2001,Green2001,Biham2002}.
It takes into account both the discrete nature of the
H atoms as well as the fluctuations. 
In the case of hydrogen recombination, 
its dynamical variables are the
probabilities $P_{\rm H}(N_{\rm H})$ 
that there are 
$N_{\rm H}$ 
atoms
on the grain
at time $t$. 
The time derivatives  
$\dot{P}_{\rm H}(N_{\rm H})$, $N_{\rm H}=0, 1, 2, \dots$
are expressed 
in terms of the adsorption, reaction and desorption terms.
The master equation provides the time evolution of 
$P_{\rm H}(N_{\rm H})$, 
$N_{\rm H}=0, 1, 2, \dots$, 
from which
the recombination rate can be calculated.
The master equation approach 
has been applied to the study of
reaction networks involving multiple species
\citep{Stantcheva2002,Stantcheva2002b}.
Since the state of the system is given by the
set of number densities of all species, the
number of equations quickly increases
with the number of reactive species.
Therefore, suitable cutoffs should be imposed in order
to keep the number of equations at a tractable level. 
A Monte Carlo approach based on the master equation was
also proposed
\cite{Charnley2001},
and applied for reaction networks involving multiple
species
\citep{Stantcheva2002b}.

In this paper we present a set of moment equations that 
can be derived from the master equation,
and use it to calculate of the rate of molecular 
hydrogen formation on small grains.
In the moment equations, 
the time derivatives
of the moments
$\langle N_{\rm H}^k \rangle$, $k=1,2,\dots$,
of the distribution
$P_{\rm H}(N_{\rm H})$
are expressed in terms of
$\langle N_{\rm H} \rangle$,
$\langle N_{\rm H}^2 \rangle$,
$\dots$,
$\langle N_{\rm H}^{k+1} \rangle$.
To simulate the moment equations
we impose a
cutoff at a suitable value of
$k$, giving rise to $k$ coupled linear differential equations.
With such cutoffs, 
the moment equations can be used efficiently to
calculate the formation rate of molecular hydrogen for both
steady state and time dependent conditions.
Furthermore, they provide a useful asymptotic expression 
for the rate of molecular hydrogen formation 
in the limit of very small grains. 
This expression can 
be used in order
to integrate the H$_2$ production rate over 
the grain-size distribution in an interstellar cloud
\cite{Mathis1977,Draine1985,Draine1985,Mathis1990,Mathis1996,Weingartner2001,Cox2002},
and obtain the total production rate per unit volume of the cloud.

In the case of complex reaction networks involving multiple
species, the number of moment equations quickly increases
as the number of reactive species grows.
Furthermore, the higher moments of the distribution are large
and cannot be neglected.
Thus, setting up the cutoffs in order to limit the number of equations
is complicated and tends to introduce significant errors.
As a result, for the case of multiple species we have not been able
to develop an effective computational framework based on the moment
equations. 
Here we present the set of moment equations for two species,
such as 
hydrogen and oxygen, where the moments take the form
$\langle N_{\rm H}^k N_{\rm O}^{\ell} \rangle$,
$k,l=0,1,2,\dots$,
and discuss the approximations that may be used in order to set suitable
cutoffs in these equations.
In spite of these difficulties, 
the moment equation approach has some advantages that justify further attempts to
develop it for the case of multiple species.
One advantage is that the moment equations 
are a direct generalization of
the rate equations and resemble their structure. 
One may thus expect that they will be suitable for incorporation 
into models of interstellar chemistry based on rate equations.
Another advantage is that in this approach, the moments, which quantify
the production rate of molecules, are directly calculated
unlike the case of the master equation where the calculation of moments
requires further processing.

The paper is organized as follows. 
In Sect. 2 we consider reactions involving a 
single atomic specie, focusing on the case of hydrogen recombination.
In Sect. 3 we consider reactions involving two atomic species,
such as hydrogen and oxygen. 
In each of these Sections we
briefly describe the rate equations and master equation for 
the system under study 
and then introduce and analyze the moment equations.
The results are discussed and summarized in
Sect. 4.
 
\section{Hydrogen recombination}

Consider a diffuse interstellar cloud dominated by 
a density
$\rho_{\rm H}$
(cm$^{-3}$)
of H atoms,
and includes some density
of dust grains.
The typical velocity 
$v_{\rm H}$ (cm s$^{-1}$)
of H atoms in the gas phase is
given by

\begin{equation}
v_{\rm H} = \sqrt{ {8 \over \pi} {k_B T_{\rm gas} \over m_H} },
\label{eq:velocity}
\end{equation}

\noindent
where $m_H=1.67 \cdot 10^{-24}$ (gram)
is the mass of an H atom
and $T_{\rm gas}$ is the gas temperature
\citep{Landau1980}.
To evaluate the flux of atoms onto grain surfaces
we will assume, 
for simplicity,
that the grains are 
spherical with a
radius $r$ (cm).
The cross section of a grain is $\sigma = \pi r^2$
and its surface area is $4 \sigma$.
The number of adsorption sites on a grain is denoted by $S$.
The flux
$F_{\rm H}$ (atoms s$^{-1}$) 
of H atoms onto the surface of a single grain 
is given  by
$F_{\rm H} = \rho_{\rm H} v_{\rm H} \sigma$.
This flux can also be expressed according to
$F_{\rm H} = f_{\rm H} \cdot S$,
where 
$f_{\rm H}$ (ML s$^{-1}$)
is given by
$f_{\rm H} = \rho_{\rm H} v_{\rm H}/(4s)$,
and $s$ (sites cm$^{-2}$) 
is the density of adsorption sites on the surface.
The H atoms stick to the surface and hop as random walkers
between adjacent sites until they either desorb as atoms
or recombine into molecules.
The desorption rate of an H atom on the surface is 

\begin{equation}
W_{\rm H} =  \nu \cdot \exp (- E_{1} / k_{B} T),  
\label{eq:P1}
\end{equation}

\noindent
where $\nu$ is the attempt rate 
(typically assumed to be $10^{12}$ s$^{-1}$), 
$E_{1}$ 
is the activation energy barrier for desorption 
of an H atom and $T$ is the surface temperature.
The hopping rate of an H atoms is

\begin{equation} 
a_{\rm H} =  \nu \cdot \exp (- E_{0} / k_{B} T), 
\label{eq:Alpha}
\end{equation}

\noindent
where
$E_{0}$ is the activation energy barrier for H diffusion. 
Here we assume that diffusion occurs only by thermal hopping,
in agreement with recent experimental results
\citep{Katz1999}.
Throughout this paper we use the parameters obtained
experimentally for amorphous carbon,
namely the activation energies are
$E_0=44.0$ meV
and
$E_1=56.7$ meV
\citep{Katz1999},
and the density of adsorption sites on the
surface is
$s \simeq 5 \times 10^{13}$ 
(sites cm$^{-2}$)
\citep{Biham2001}.
For the
density of hydrogen atoms in the gas phase
we take
$\rho_{\rm H}=10$ 
(atoms cm$^{-3}$).
The gas temperature is taken as 
$T_{\rm gas}=90$ K,
thus 
$v_H=1.37 \times 10^5$ (cm s$^{-1}$). 
These are typical values for diffuse interstellar clouds.

The number of H atoms on the grain is 
denoted by
$N_{\rm H}$
and its expectation value
under the given conditions is denoted by
$\langle N_{\rm H} \rangle$.
The  rate 
$A_{\rm H} = a_{\rm H}/S$
is approximately the inverse of the time
$t_s$
required for an atom
to visit nearly all the
adsorption sites on the grain surface.
This is due to the fact that in two dimensions the 
number of distinct sites visited by a random walker
is linearly proportional to the number of steps, up
to a logarithmic correction
\citep{Montroll1965}.

\subsection{Rate equation}

The rate equation describing
H$_2$ formation on dust grain surfaces takes the form

\begin{equation}
\label{eq:Ngrain}
{ {d{ \langle N_{\rm H} \rangle }} \over {dt}}  =  F_{\rm H} 
- W_{\rm H} \langle N_{\rm H} \rangle - 2 A_{\rm H} {\langle N_{\rm H} \rangle}^{2}. 
\label{eq:N1grain} \\
\end{equation}

\noindent
The first term on the right hand side describes the flux of H atoms,
the second term describes the desorption of H atoms and the third
term describes the diffusion and recombination. 
Here we assume, for simplicity, that all H$_2$ molecules dosorb from the
surface upon formation.
The production rate 
$R_{\rm H_2}^{\rm grain}$ 
(molecules s$^{-1}$)
of H$_2$ molecules from a single grain is given by

\begin{equation}
R_{\rm H_2}^{\rm grain} = A_{\rm H} \langle N_{\rm H} \rangle^2.
\label{eq:RH2}
\end{equation}

\noindent
As long as
$\langle N_{\rm H} \rangle \gg 1$
Eq. 
(\ref{eq:RH2})
provides a very good evaluation of the H$_2$ formation rate.
However, for 
very small grains and low flux, for which
$\langle N_{\rm H} \rangle \simeq 1$
or less, this mean-field approach fails,
and tends to
overestimate the H$_2$ production. 
This is because the rate equation does not take into
account the discrete nature of the 
population of
H atoms, 
and the fact that it takes
at least two
atoms on the surface simultaneously to enable recombination.
In this limit, the master equation approach is needed in order
to evaluate the H$_2$ formation rate.

\subsection{Master equation}
\label{sec:MasterModel}

We will now describe the
master equation for H$_2$ formation 
on small interstellar grains,
exposed to a flux $F_{\rm H}$ of H atoms.
At any given time the number of H atoms adsorbed on the grain
may be $N_{\rm H}=0, 1, 2, \dots$,
and the
probability that there are $N_{\rm H}$ hydrogen atoms  
on a grain 
is given by
$P_{\rm H}(N_{\rm H})$,
where

\begin{equation}
\sum_{N_{\rm H}=0}^{\infty} P_{\rm H}(N_{\rm H}) =1.
\label{eq:normalization}
\end{equation}

\noindent
The time 
derivatives
of these probabilities,
$\dot P_{\rm H}(N_{\rm H})$,
are given by 

\begin{eqnarray}
\label{eq:Nmicro}
\dot P_{\rm H}(0) &=& - F_{\rm H} P_{\rm H}(0) + W_{\rm H} P_{\rm H}(1) 
+ 2 \cdot 1 \cdot A_{\rm H} P_{\rm H}(2) \nonumber \\
\dot P_{\rm H}(1) &=&  F_{\rm H} \left[ P_{\rm H}(0) - P_{\rm H}(1) \right] 
\nonumber\\&+& W_{\rm H} \left[ 2 P_{\rm H}(2) - P_{\rm H}(1) \right] 
                + 3 \cdot 2 \cdot A_{\rm H} P_{\rm H}(3) \nonumber \\
&\vdots& \nonumber \\
\dot P_{\rm H}(N_{\rm H}) &=&  
F_{\rm H} \left[ P_{\rm H}(N_{\rm H}-1) - P_{\rm H}(N_{\rm H}) \right] 
\nonumber\\
&+& W_{\rm H} \left[ (N_{\rm H}+1) P_{\rm H}(N_{\rm H}+1) - N_{\rm H} P_{\rm H}(N_{\rm H}) \right] \nonumber \\
                &+& A_{\rm H} [ (N_{\rm H}+2)(N_{\rm H}+1) P_{\rm H}(N_{\rm H}+2) \nonumber\\
&-&  N_{\rm H}(N_{\rm H}-1) P_{\rm H}(N_{\rm H}) ].   \\
&\vdots& \nonumber 
\end{eqnarray}

\noindent
Each of these equations includes three terms.
The first term describes the effect of the incoming flux $F_H$ on the probabilities.
The probability $P_{\rm H}(N_{\rm H})$ increases when an H atom is adsorbed on a grain that already
has $N_{\rm H}-1$ adsorbed atoms 
[at a rate of $F_{\rm H} P_{\rm H}(N_{\rm H}-1)$], 
and decreases when a new atom is adsorbed on a grain with
$N_{\rm H}$ atoms on it
[at a rate of $F_{\rm H} P_{\rm H}(N_{\rm H})$].
The second term includes the effect of desorption. 
An H atom that is
desorbed from a grain with $N_{\rm H}$ atoms,  decreases the
probability $P_{\rm H}(N_{\rm H})$
[at a rate of
$N_{\rm H} W_{\rm H} P_{\rm H}(N_{\rm H})$], 
and increases the probability
$P_{\rm H}(N_{\rm H}-1)$
at the same rate.
The third term describes the effect of recombination on the number of adsrobed
H atoms. 
The production of one molecule reduces this number from $N_{\rm H}$ to $N_{\rm H}-2$.
For one pair of H atoms the recombination rate is 
proportional to 
the sweeping rate
$A_{\rm H}$ 
multiplied by 2 since both atoms are mobile
simultaneously.
This rate is multiplied by
the number of possible pairs of atoms, namely
$N_{\rm H}(N_{\rm H}-1)/2$. 
Note that the equations for 
$\dot P_{\rm H}(0)$ 
and
$\dot P_{\rm H}(1)$ 
do not include all the terms, because at least one H 
atom is required for desorption to occur and at least two
for recombination.
The rate of formation of H$_2$ molecules,
$R_{\rm H_2}^{\rm grain}$ (molecules s$^{-1}$), 
on the surface of a single grain is thus
given by 



\begin{equation}
R_{\rm H_2}^{\rm grain} = 
A_{\rm H} [\langle N_{\rm H}^2\rangle - \langle N_{\rm H}\rangle]
\label{eq:Rgrain}
\end{equation}

\noindent
where

\begin{equation}
\langle N_{\rm H}^k \rangle = \sum_{N_{\rm H}=0}^{\infty} N_{\rm H}^k P_{\rm H}(N_{\rm H}).
\label{eq:defmomentk}
\end{equation}

\noindent
is the $k$th moment of the distribution.

The time dependence of 
$R_{\rm H_2}^{\rm grain}$
can be obtained by numerically integrating Eqs.
(\ref{eq:Nmicro})
using a standard Runge-Kutta stepper.
For the case of steady state,
namely
$\dot P_{\rm H}(N_{\rm H})=0$
for all $N_{\rm H}$,
an analytical solution for 
$P_{\rm H}(N_{\rm H})$
is available,
given in terms of
$A_{\rm H}/W_{\rm H}$
and
$W_{\rm H}/F_{\rm H}$
\citep{Green2001,Biham2002}.

\subsection{Moment equations}
\label{sec:Moment} 

By expanding the time derivatives
$\langle \dot N_{\rm H}^k\rangle$, $k=1,2,\dots$,
using Eq.
(\ref{eq:defmomentk})
and inserting the expression for
$\dot P_{\rm H} (N_{\rm H})$
from the master equation
(\ref{eq:Nmicro}),
one obtains the moment equations:

\begin{eqnarray}
{d \langle N_{\rm H} \rangle \over {dt} } &=& F_{\rm H} 
+( - W_{\rm H} + 2A_{\rm H})  \langle N_{\rm H} \rangle 
- 2 A_{\rm H} \langle N_{\rm H}^2 \rangle  \nonumber \\
{d \langle N_{\rm H}^2 \rangle \over {dt} } &=& F_{\rm H} 
+( 2 F_{\rm H}+ W_{\rm H}-4A_{\rm H} ) \langle N_{\rm H} \rangle 
\nonumber\\
&+& (8 A_{\rm H} - 2 W_{\rm H}) \langle N_{\rm H}^2 \rangle
- 4 A_{\rm H} \langle N_{\rm H}^3 \rangle \nonumber \\
{d \langle N_{\rm H}^3 \rangle \over {dt} } &=& F_{\rm H} 
+( 3 F_{\rm H}- W_{\rm H}+8A_{\rm H} ) \langle N_{\rm H} \rangle
\nonumber\\
&+& (3 F_{\rm H} + 3 W_{\rm H} - 20 A_{\rm H} ) \langle N_{\rm H}^2 \rangle 
\nonumber \\
&+& (18 A_{\rm H} - 3 W_{\rm H}) \langle N_{\rm H}^3 \rangle 
- 6 A_{\rm H} \langle N_{\rm H}^4 \rangle \nonumber \\
\vdots \nonumber \\
{d \langle N_{\rm H}^k \rangle \over {dt} } &=& 
F_{\rm H} \langle (1+N_{\rm H})^k - N_{\rm H}^k \rangle 
\nonumber\\
&+& W_{\rm H} \langle N_{\rm H} [(N_{\rm H}-1)^k - N_{\rm H}^k] \rangle \nonumber \\
&+& A_{\rm H} \langle  N_{\rm H}(N_{\rm H}-1)[(N_{\rm H}-2)^k - N_{\rm H}^k] \rangle.\nonumber \\
\vdots 
\label{eq:moments}
\end{eqnarray}

\noindent
This is a set of coupled differential equations,
that are
linear in the moments
$\langle N_{\rm H}^k \rangle$.
The equation for the $k$th moment depends on all the moments
up the the $(k+1)$th order.
A numerical integration of these equations would provide the
values of these moments at any given time. The H$_2$ production
rate, that depends only on the first and second moments, could
then be obtained.
However, the difficulty is that we need to truncate this set of equations,
say after the equation of the $k$th moment.
The $(k+1)$th moment appears in this equation and we end up with more
unknowns than equations.
The $(k+1)$th moment should then be approximated as a function of the
first $k$ moments.
The approximation should be examined carefully since 
$\langle N_{\rm H}^{k+1} \rangle$
is not a small quantity
(unlike the case when the master equation is truncated).
In fact, due to the discreteness of 
$N_{\rm H}$,
the moments increase monotonically with their order, namely
$\langle N_{\rm H}^{k} \rangle \le \langle N_{\rm H}^{k+1} \rangle$,
$k=1,2,\dots$ 
(and the $\le$ symbol is replaced by the $<$ symbol if 
$P_{\rm H}(N_{\rm H})>0$ for at least one value of $N_{\rm H}>1$).

\subsubsection{Setting the cutoffs}

Consider the case in which only 
up to $k$ hydrogen atoms are allowed to reside simultaneously
on the surface of a grain.
The moments of the resulting truncated distribution
$P_{\rm H}(N_{\rm H})$
will be

\begin{eqnarray}
\langle N_{\rm H}^{1} \rangle &=& P_{\rm H}(1) + 2 P_{\rm H}(2)+\dots+kP_{\rm H}(k) \nonumber \\
\langle N_{\rm H}^{2} \rangle &=& P_{\rm H}(1) + 4 P_{\rm H}(2)+\dots+k^2 P_{\rm H}(k) \nonumber \\
\vdots \nonumber \\
\langle N_{\rm H}^{k} \rangle &=& P_{\rm H}(1) + 2^k P_{\rm H}(2)+\dots + k^k P_{\rm H}(k) \nonumber \\ 
\langle N_{\rm H}^{k+1} \rangle &=& P_{\rm H}(1) + 2^{k+1} P_{\rm H}(2)+\dots 
+ k^{k+1} P_{\rm H}(k). 
\label{eq:cutoff}
\end{eqnarray}

\noindent
The first $k$ equations in
Eqs. 
(\ref{eq:cutoff})
can be used in order to express
$P_{\rm H}(1)$ ,$\dots$, $P_{\rm H}(k)$
in terms of
$\langle N_{\rm H}^{1} \rangle$, $\dots$, $\langle N_{\rm H}^{k} \rangle$.
The results can then be plugged into the $(k+1)$th equation
in Eqs.
(\ref{eq:moments}), 
thus expressing
$\langle N_{\rm H}^{k+1} \rangle$
in terms of the first $k$ moments.
For example, in the case of $k=2$ 
we obtain the third moment
as a function of the first two:

\begin{eqnarray}
\langle N_{\rm H}^{3} \rangle =  
3 \langle N_{\rm H}^{2} \rangle
- 2 \langle N_{\rm H}^{1} \rangle.
\label{eq:cutoff3}
\end{eqnarray}

\noindent
Similarly, when
$P_{\rm H}(N_{\rm H})$ 
is truncated after $k=3$ or $k=4$ we
obtain

\begin{equation}
\langle N_{\rm H}^{4} \rangle = 
6 \langle N_{\rm H}^{1} \rangle    
-11 \langle N_{\rm H}^{2} \rangle    
+ 6 \langle N_{\rm H}^{3} \rangle 
\end{equation}

\noindent
and

\begin{equation}
\langle N_{\rm H}^{5} \rangle = 
-24 \langle N_{\rm H}^{1} \rangle    
+50 \langle N_{\rm H}^{2} \rangle    
-35 \langle N_{\rm H}^{3} \rangle    
+10 \langle N_{\rm H}^{4} \rangle, 
\end{equation}

\noindent
respectively. 
In general, when up to $k$ hydrogen atoms are allowed to reside on a grain
the $(k+1)$th moment,
$\langle N_{\rm H}^{k+1} \rangle$,
can be expressed as a linear combination of 
the first $k$ moments according to

\begin{equation}
\langle N_{\rm H}^{k+1} \rangle=\sum_{n=1}^k 
C_{k+1}(n) \langle N_{\rm H}^{n} \rangle . 
\label{eq:lincomb}
\end{equation}

\noindent
The coefficients $C_{k+1}(n)$ are obtained from the solution of 
a set of linear algebraic equations, that can be
expressed in a matrix form as 

\begin{equation}
V \vec{C} = \vec{v}.
\end{equation}

\noindent
The matrix $V$ is the Vandermonde matrix of size $k \times k$,  
namely,
$V_{mn}=m^n$, 
where 
$m,n=1,2,\dots,k$,
and
the vector
$\vec v$
consists of 
$v_m=m^{k+1}$,
$m=1,2,\dots,k$.    
The 
desired coefficients are the elements of the
vector $\vec{C}= [C_{k+1}(1),\dots,C_{k+1}(k)]$. 
The coefficient 
$C_{k+1}(n)$
turn out to be equal to
the coefficient of $x^n$   
in the polynomial
\cite{Bender2002} 

\begin{equation}
Q_k(x)=-\prod_{j=0}^k(x-j).
\end{equation}

\noindent
Having an expression for
$\langle N_{\rm H}^{k+1} \rangle$
as a linear combination of the 
first $k$ moments, we can now insert it
into the last equation in
Eq.
(\ref{eq:moments}).
We then obtain a set of $k$ coupled linear differential
equations for the first $k$ moments.
This set of equations can be simulated using a standard Runge-Kutta
or any other stepper routine.
A convenient choice of initial condition may be the case of an empty grain.
In this case
the moment equations are initiated with
$\langle N_{\rm H}^n \rangle =0$ 
for all
$n \ge 1$ 
(while the master equation is initiated with 
$P_{\rm H}(0)=1$ 
and 
$P_{\rm H}(n)=0$ for $n \ge 1$).

\subsubsection{Calculations and results}

To determine the desirable value of the cutoff $k$,
for steady state calculations using the moment equations, 
it is useful to 
first calculate 
$\langle N_{\rm H} \rangle$
using
the rate equations 
[Eq. (\ref{eq:N1grain})]
for the same parameters used in the moment equations. 
We found that a suitable choice for the truncation is
$k =\lceil \langle N_{\rm H} \rangle + C \rceil$
where 
$\lceil x \rceil$
is the smallest integer which is larger than $x$.
The parameter $C$ is determined such that a sufficient
number of equations are included to provide a good
agreement with the master equation.
In all the calculations presented here we used
$C=1.2$.

We have used the moment equations to simulate the hydrogen recombination
process on a grain under steady state conditions, and compared the 
results with those of the rate equations and the master equation.
The surface parameters used are of amorphous carbon. 
The gas density and temperature are those specified above,
and the surface temperature is $T=18$ K.

\begin{figure}
\newsavebox{\figaa}
\epsfxsize=6.7cm
\sbox{\figaa}{\epsffile{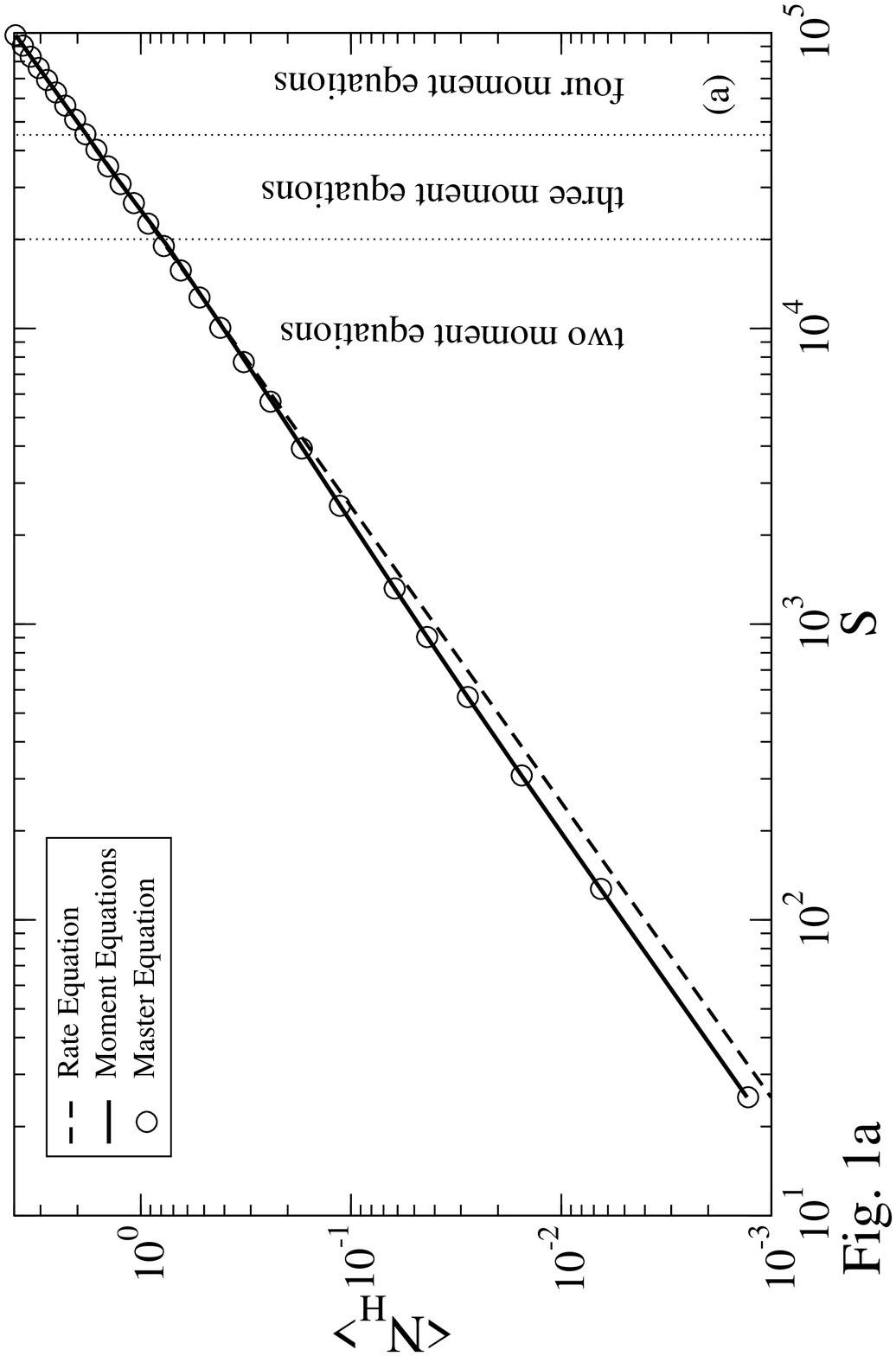}}
\rotr\figaa
\newsavebox{\figab}
\epsfxsize=6.7cm
\sbox{\figab}{\epsffile{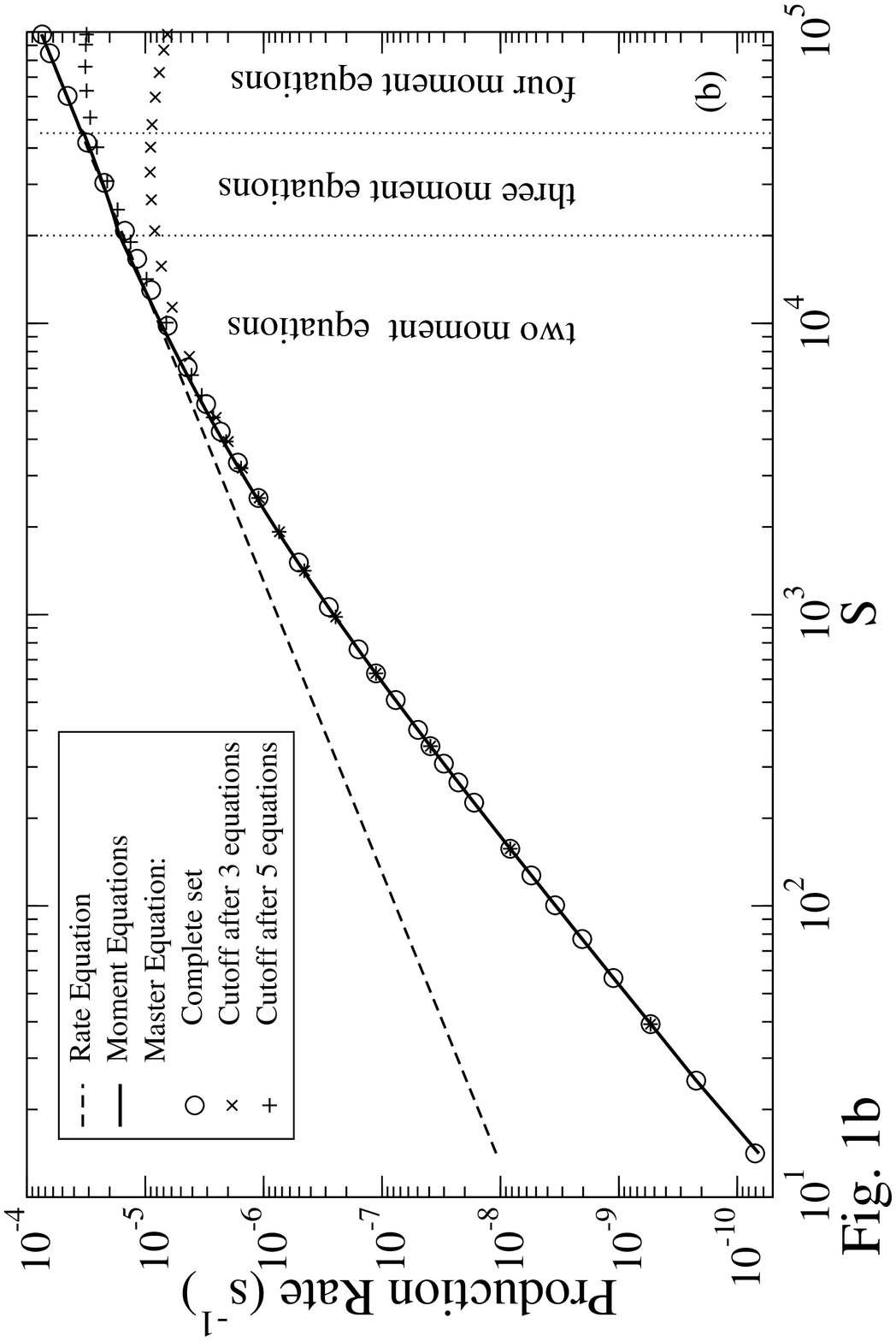}}
\rotr\figab
\caption{
(a) The expectation value of the 
number of H atoms
on the grain as a function of the number
of adsorption sites, $S$, on the grain, obtained from the 
moment equations, the master equation and the rate equations;
(b) The production rate $R_{\rm H_2}^{\rm grain}$ of molecular hydrogen on a single
grain of $S$ surface adsorption sites.
The results of the moment equations (solid line) are in perfect
agreement with those of the master equation ($\bigcirc$),
while the rate equations (dashed line) over-estimate the production
rate for small grains.
The number of moment equations used for each range of grain sizes
is specified.
\label{fig:1}
}
\end{figure}

The number of H atoms on the grain is shown in Fig. 
\ref{fig:1}(a)
as a function of the number of adsorption sites, $S$, 
using the rate equations (dashed line),
master equation ($\bigcirc$)
and 
moment equations (solid line).
The production rate of H$_2$ molecules on the surface of 
a single grain is shown in 
Fig. 
\ref{fig:1}(b)
as a function of $S$.
The results of the rate equations (dashed line) are simply
linearly proportional to $S$. 
For large enough grains the results of the 
rate equations coincide with those of the 
master equation ($\bigcirc$)
and the moments equations (solid line).
For small grains, the results of the moment equations 
are in perfect agreement with the 
master equation,
while the rate equations greatly over-estimate the production rate.
In the simulations of the moment equations, the cutoff was determined 
according the the procedure specified above. 
The number of equations used in each range of grain sizes is specified.
Note that using this procedure we obtain
a perfect agreement between the moment equations
and the master equation for the entire range of grain sizes.
Results are also shown for the master equation with cutoffs 
after the third equation ($\times$) and after the fifth equation
($+$).
It is shown that while the sets of two, three and four moment equations 
provide a perfect agreement with the complete master equation in the
specified domains, the master equation with cutoffs (including at least
as many equations) exhibits significant deviations.

In the limit of very small grains, only two moment equations
are needed in order to obtain perfect agreement with the 
master equation. 
These equation
are derived from
the first two moment equations
in Eqs.
(\ref{eq:moments})
and the cutoff condition
given in Eq.
(\ref{eq:cutoff3}).
They take the form

\begin{eqnarray}
{ d \langle  N_{\rm H} \rangle \over {dt} } &=& 
F_{\rm H} + (2 A_{\rm H} - W_{\rm H}) \langle  N_{\rm H} \rangle 
- 2 A_{\rm H} \langle  N_{\rm H}^2 \rangle \nonumber \\
{ d \langle  N_{\rm H}^2 \rangle \over {dt} } &=& 
F_{\rm H} + (2 F_{\rm H} + W_{\rm H} + 4 A_{\rm H}) \langle  N_{\rm H} \rangle 
- (4 A_{\rm H} + 2 W_{\rm H}) \langle  N_{\rm H}^2 \rangle. 
\label{eq:mom2eq}
\end{eqnarray}

\begin{figure}
\newsavebox{\figba}
\epsfxsize=6.7cm
\sbox{\figba}{\epsffile{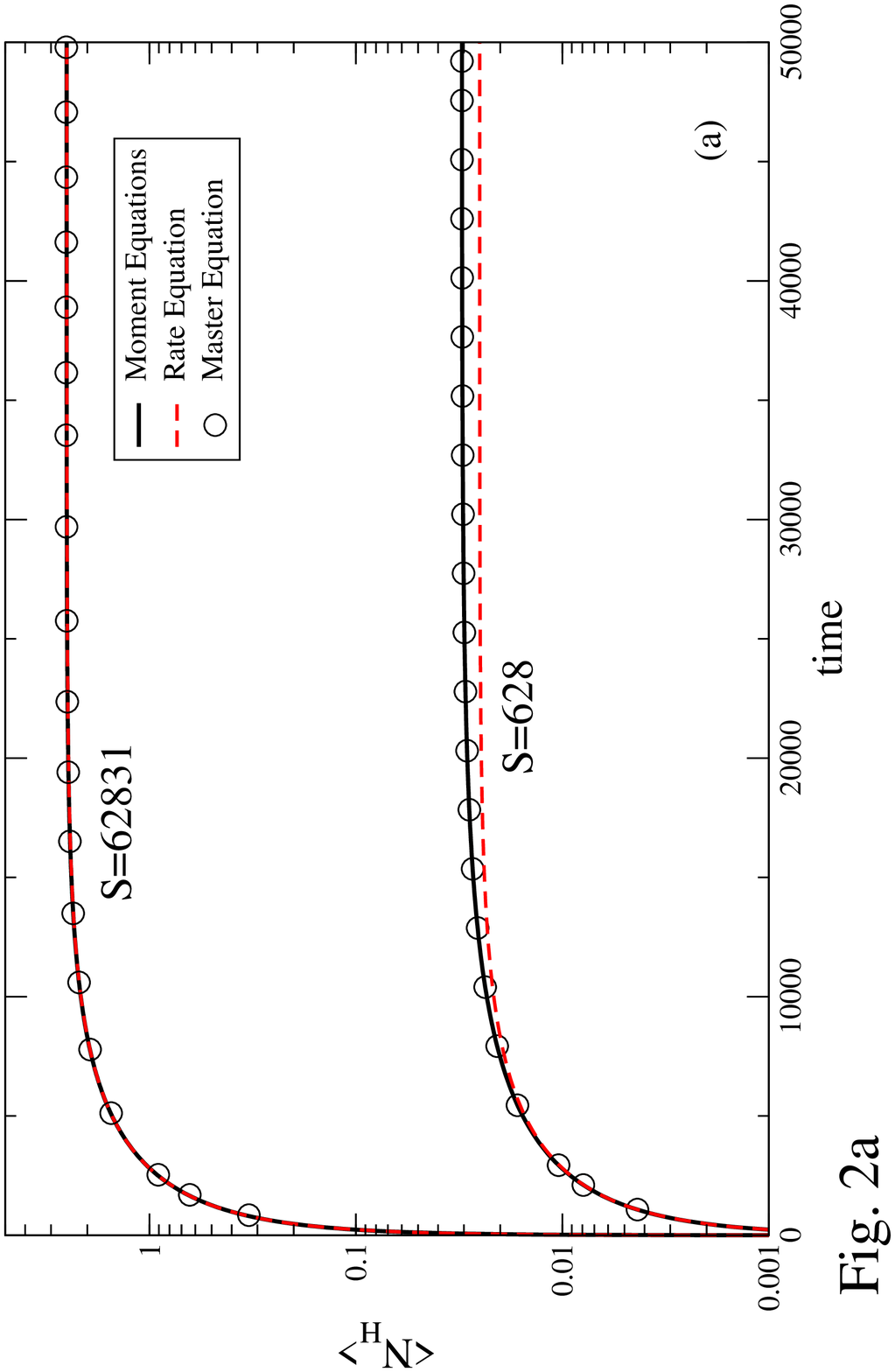}}
\rotr\figba
\newsavebox{\figbb}
\epsfxsize=6.7cm
\sbox{\figbb}{\epsffile{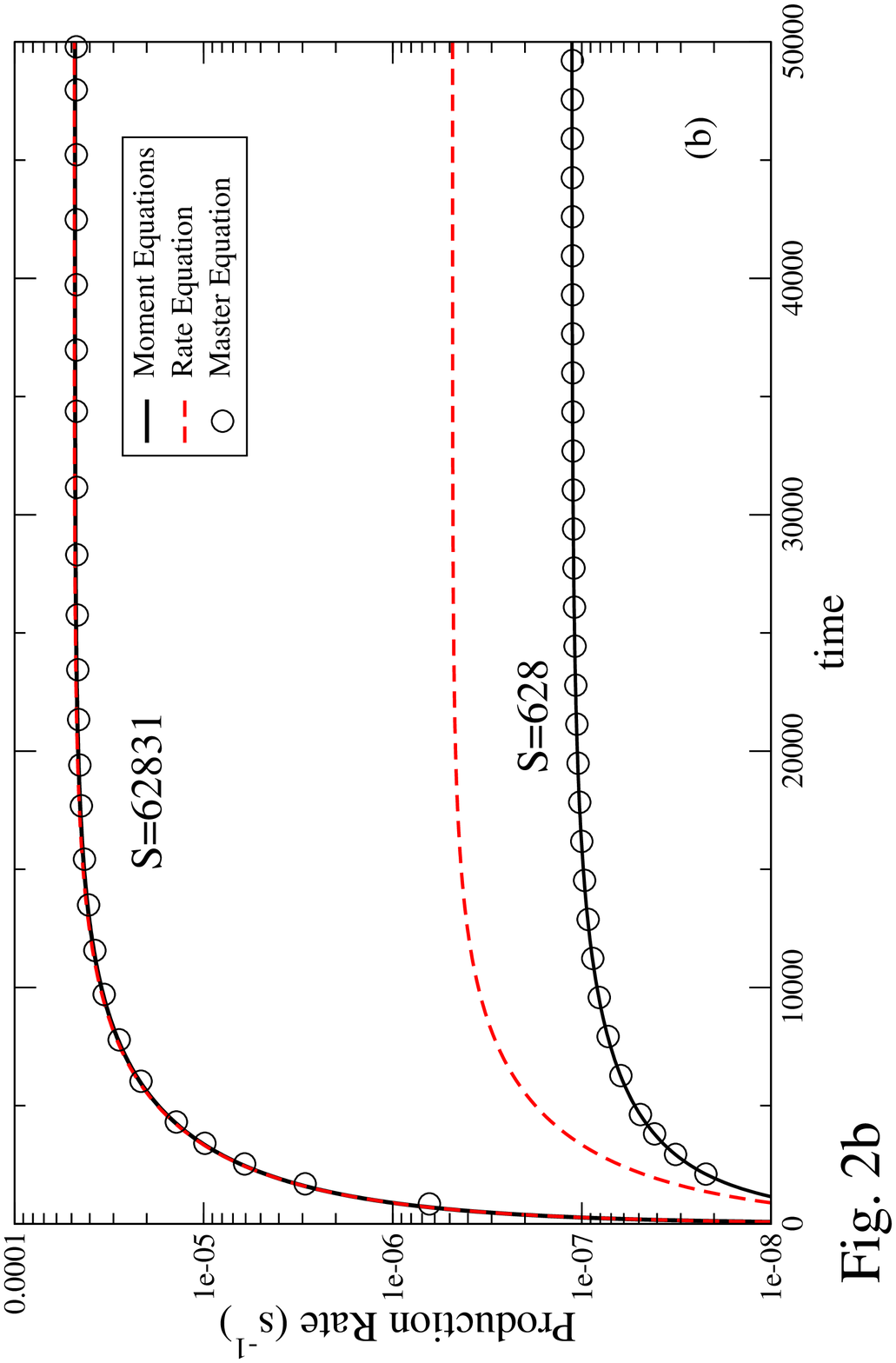}}
\rotr\figbb
\caption{
(a) The expectation value for the number of H atoms on the grain
vs. time; and (b)
the production rate 
$R_{\rm H_2}^{\rm grain}$ of molecular hydrogen vs. time,
for a large grain ($S=62,831$ sites)
and for a small grain ($S=628$ sites).
The initial condition is of an empty
grain surface, while the surface parameters, the temperature
and the flux are identical to those used in Fig. 1.
For the large grain, it is found that the moment equations (solid line),
master equation ($\bigcirc$) 
and the rate equations (dashed line) are all in perfect
agreement even during the transient time.
For the small grain, the set of two moment equations
is in perfect agreement with the complete master equation,
while the rate equations deviate significantly.
\label{fig:2}
}
\end{figure}

\noindent
We find that these equations are dynamically stable and
can be used for time-dependent simulations.
In Fig. 2 we present the results of time-dependent simulations
using Eqs.
(\ref{eq:mom2eq})
for a large grain ($S=62,831$ sites)
and for a small grain ($S=628$ sites).
The expectation value 
$\langle N_{\rm H}\rangle$
vs. time is shown in Fig. 2(a),
and the production rate 
$R_{\rm H_2}^{\rm grain}$
of molecular hydrogen is 
shown in Fig.
2(b).
The initial condition is of an empty
grain surface, while the surface parameters, the temperature
and the flux are identical to those used in Fig. 1.
For the large grain, it is found that the moment equations,
master equation and the rate equation are all in perfect
agreement even during the transient time.
For the small grain, the set of two moment equations
is in perfect agreement with the complete master equation,
while the rate equation deviates significantly.
For steady state conditions, Eqs.
(\ref{eq:mom2eq})
can be solved exactly, giving rise to

\begin{eqnarray}
\langle  N_{\rm H} \rangle &=&
{ F_{\rm H}(A_{\rm H}+W_{\rm H}) \over {2A_{\rm H}F_{\rm H} 
+ W_{\rm H}A_{\rm H} + W_{\rm H}^2} } 
\nonumber \\
\langle  N_{\rm H}^2 \rangle &=&
{ F_{\rm H}(F_{\rm H}
+A_{\rm H}+W_{\rm H}) \over {2A_{\rm H}F_{\rm H} 
+ W_{\rm H}A_{\rm H} + W_{\rm H}^2} }.  
\label{eq:mom2exsol}
\end{eqnarray}

\noindent
A simple exact expression for the production rate 
$R_{\rm H_2}^{\rm grain}$, of molecular hydrogen, given by Eq.
(\ref{eq:Rgrain}),
can now be obtained for steady state conditions
in the limit of small grains and low flux. 
It takes the form

\begin{equation}
R_{\rm H_2}^{\rm grain} = {f_{\rm H}^2 \over W_{\rm H}} 
\left[{S^2  \over {1 + {S \over {a_{\rm H}/W_{\rm H}} } 
+ 2 {S \over {W_{\rm H}/f_{\rm H}} } } }\right].
\label{eq:Rasympt}
\end{equation}

\noindent
We observe that, as long as 
$S$ is larger than both
$s_{\rm visit}=a_{\rm H}/W_{\rm H}$ 
(the number of sites that an H atom typically visits before it desorbs)
and
$s_{\rm vacant}=W_{\rm H}/f_{\rm H}$ 
(the typical number of vacant sites per adsorbed H atom),
the production rate
is linearly proportional to $S$, and the grain size does not play any special role
\cite{Biham2002}.
In the case of small grains for which $S$ is smaller than both 
$s_{\rm visit}$ 
and
$s_{\rm vacant}$ 
the production rate is reduced and becomes proportional to $S^2$.
The production efficiency
$\eta(S) = 2 R_{\rm H_2}^{\rm grain} / F_{\rm H}$
then takes the form

\begin{equation}
\eta(S) = {2 f_{\rm H} \over W_{\rm H}} 
\left[{S  \over {1 + {S \over {a_{\rm H}/W_{\rm H}} } + 
2 {S \over {W_{\rm H}/f_{\rm H}} } } }\right].
\label{eq:eta_asympt}
\end{equation}

\begin{figure}
\newsavebox{\figc}
\epsfxsize=6.7cm
\sbox{\figc}{\epsffile{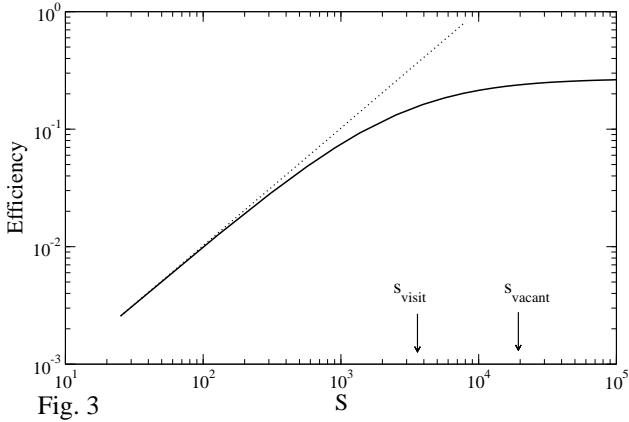}}
\rotr\figc
\caption{
The efficiency $\eta(S)$ of molecular hydrogen formation
on a grain of $S$ assorption sites is presented. It is
shown that in the limit of small grains the efficiency
depends linearly on $S$, while for large grains it saturates
towards the efficiency predicted by the rate equations and
becomes independent of $S$.
\label{fig:3}
}
\end{figure}

\noindent
In Fig. 3 we present the production efficiency,
given by Eq.
(\ref{eq:eta_asympt}),
as a function of $S$ under the same conditions as those used
in Fig. 1.
For very small grains, namely for
$S < \min \{ s_{\rm visit},s_{\rm vacant} \}$,
the efficiency $\eta(S)$ 
is linearly proportional to 
$S$, 
while for 
$S > \max \{ s_{\rm visit},s_{\rm vacant} \}$
it saturates and coincides with the rate equation result that
is independent of $S$.

\begin{figure}
\newsavebox{\figd}
\epsfxsize=6.7cm
\sbox{\figd}{\epsffile{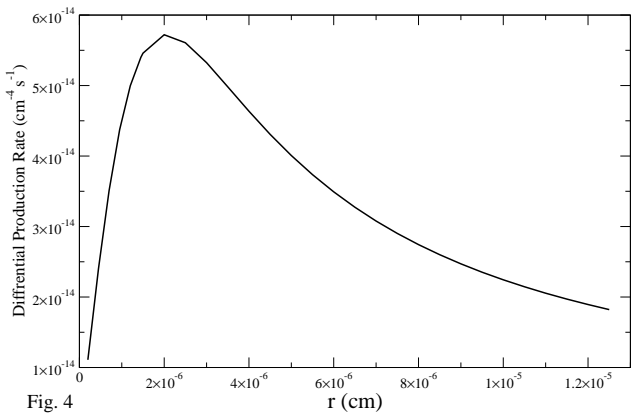}}
\rotr\figd
\caption{
The differential production rate of molecular hydrogen, 
given by the integrand in
Eq.
(\ref{eq:integ.prod.rate})
vs. the grain radius, $r$.
The largest contribution comes from very small grains, of sizes
around 20 nm. For larger grains the contribution decreases due to
the fact that their total surface area is smaller. For smaller 
grains the contribution decreases due to the reduction in the recombination
efficiency. 
\label{fig:4}
}
\end{figure}

These results can now be used in order to evaluate the total
production of molecular hydrogen per unit volume of the cloud 
from grains of all sizes.
Consider a cloud in which the grain size distribution 
is given by 
$\rho_{\rm gr}(r)$ (cm$^{-4}$)
in the range 
$r_{\rm min} < r < r_{\rm max}$,  
where 
$r_{\rm min}$  
and
$r_{\rm max}$
are the lower and upper cutoffs of the distribution
[here $\rho_{\rm gr}(r) dr$ (cm$^{-3}$) is the number density
of grains of sizes in the range (r,r+dr)].
The formation rate 
$R_{\rm H_2}$ (cm$^{-3}$ s$^{-1}$ )
of molecular hydrogen in the cloud
is given by

\begin{equation}
R_{\rm H_2} = \rho_{\rm H} v_{\rm H} \int_{ r_{\rm min} }^{ r_{\rm max} } 
\pi r^2 \rho_{\rm gr}(r) \eta(S) dr.
\label{eq:integ.prod.rate}
\end{equation}

\noindent
where $\eta(S)$ is given 
by Eq.
(\ref{eq:eta_asympt})
and
$S = 4\pi r^2 s$.

To demonstrate these ideas we will consider a simple
example in which
the grain size distribution exhibits a power-law
behavior of the form

\begin{equation}
\rho_{\rm gr}(r) = { K \over r^{\alpha}}, 
\label{eq:sizedist}
\end{equation}

\noindent
where $\alpha=3$.
The special property of this distribution is that the total
grain mass is equally distributed 
in the entire range of grain sizes
$r_{min} \le r \le r_{max}$.
Unlike the mass, 
the total surface area of all grains in the size range 
between $r$ and $r+dr$, scales like $r^{-1}$,
namely most of the grain surface area is on the small grains. 
The prefactor $K$ is determined by the 
condition that the total number density of grains
is smaller than $\rho_{\rm H}$ by a factor of $10^{-12}$.
Under this condition
$K=  2 \cdot 10^{-11} / (r_{min}^{-2} - r_{max}^{-2})$.
In this case the integral can be easily evaluated and the
result is

\begin{equation}
R_{\rm H_2} = 
{ {4 \pi^2 K \rho_{\rm H} v_{\rm H} f_{\rm H} s } 
\over 
 W_{\rm H} B
}
\ln { 
\left( 
{ {B r_{\rm max}^2 + 1} 
\over
{B r_{\rm min}^2 + 1} } 
\right)
}
\end{equation}

\noindent
where
$B = 4 \pi s [W_{\rm H}/a_{\rm H} + 2f_{\rm H}/W_{\rm H}]$.
To evaluate the contribution of grains of different sizes to the
total production rate of molecular hydrogen, we show in Fig. 4 the
differential production rate, given by the integrand in
Eq.
(\ref{eq:integ.prod.rate}),
vs. the grain radius, $r$,
where
$r_{\rm min} = 2 \times 10^{-7}$ (cm)
and
$r_{\rm max} = 1.25 \times 10^{-5}$ (cm).
Starting from the limit of large grains, the differential production
increases with decreasing grain size due to the larger 
total surface area of smaller grains. However, below some grain size
the recombination efficiency decreases and the differential production
rate starts to decrease. We observe that under these conditions the
largest contribution comes from extremely small 
grains of sizes around 20 nm.
Observation indicate that the exponent 
in Eq. 
(\ref{eq:sizedist})
is 
$\alpha = 3.5$
(see e.g. Weingartner \& Draine 2001).
Therefore, the contribution of very small grains to 
molecular hydrogen formation may be even more dominant
than in the analysis shown here.

\begin{figure}
\newsavebox{\figea}
\epsfxsize=6.7cm
\sbox{\figea}{\epsffile{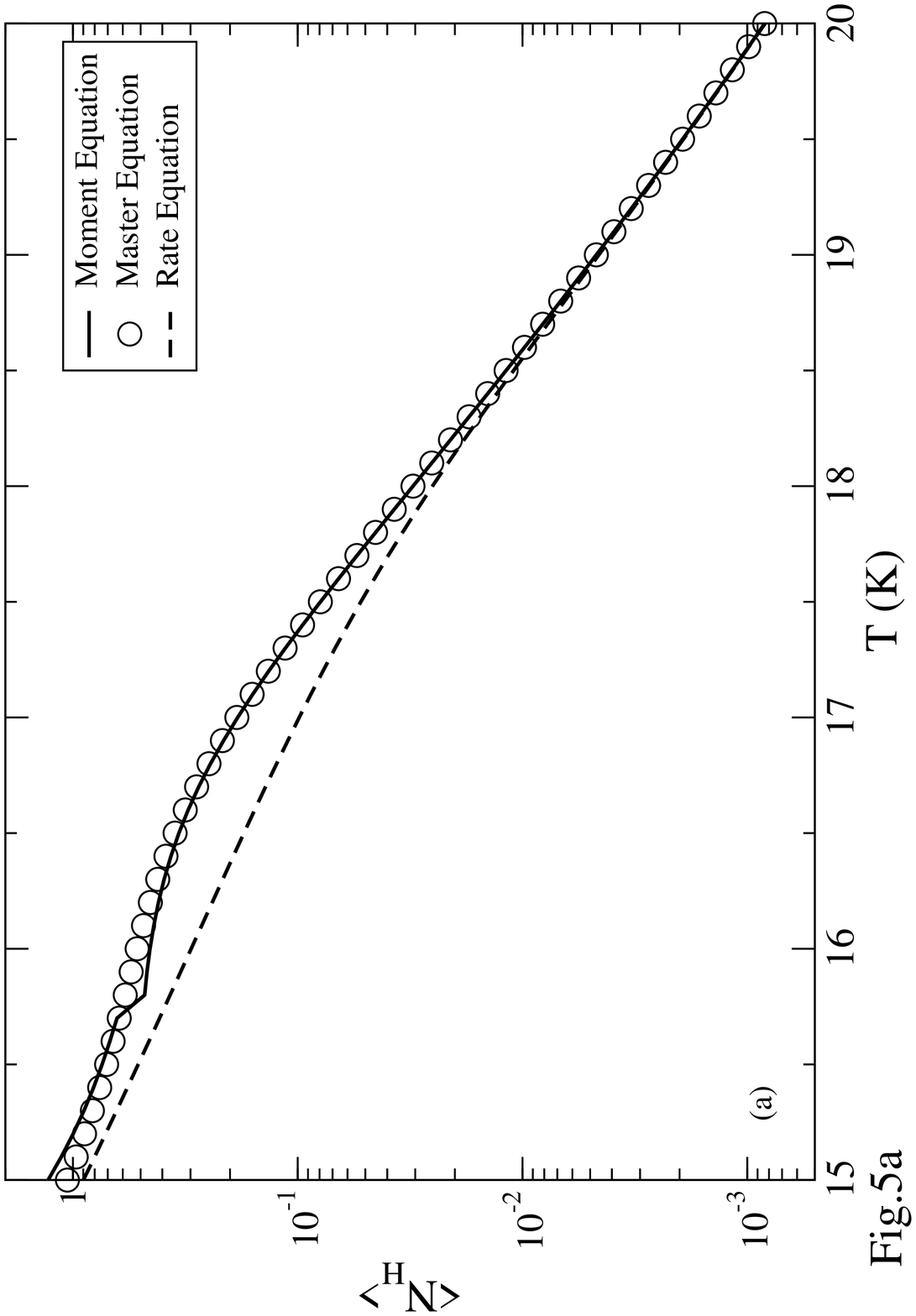}}
\rotr\figea
\newsavebox{\figeb}
\epsfxsize=6.7cm
\sbox{\figeb}{\epsffile{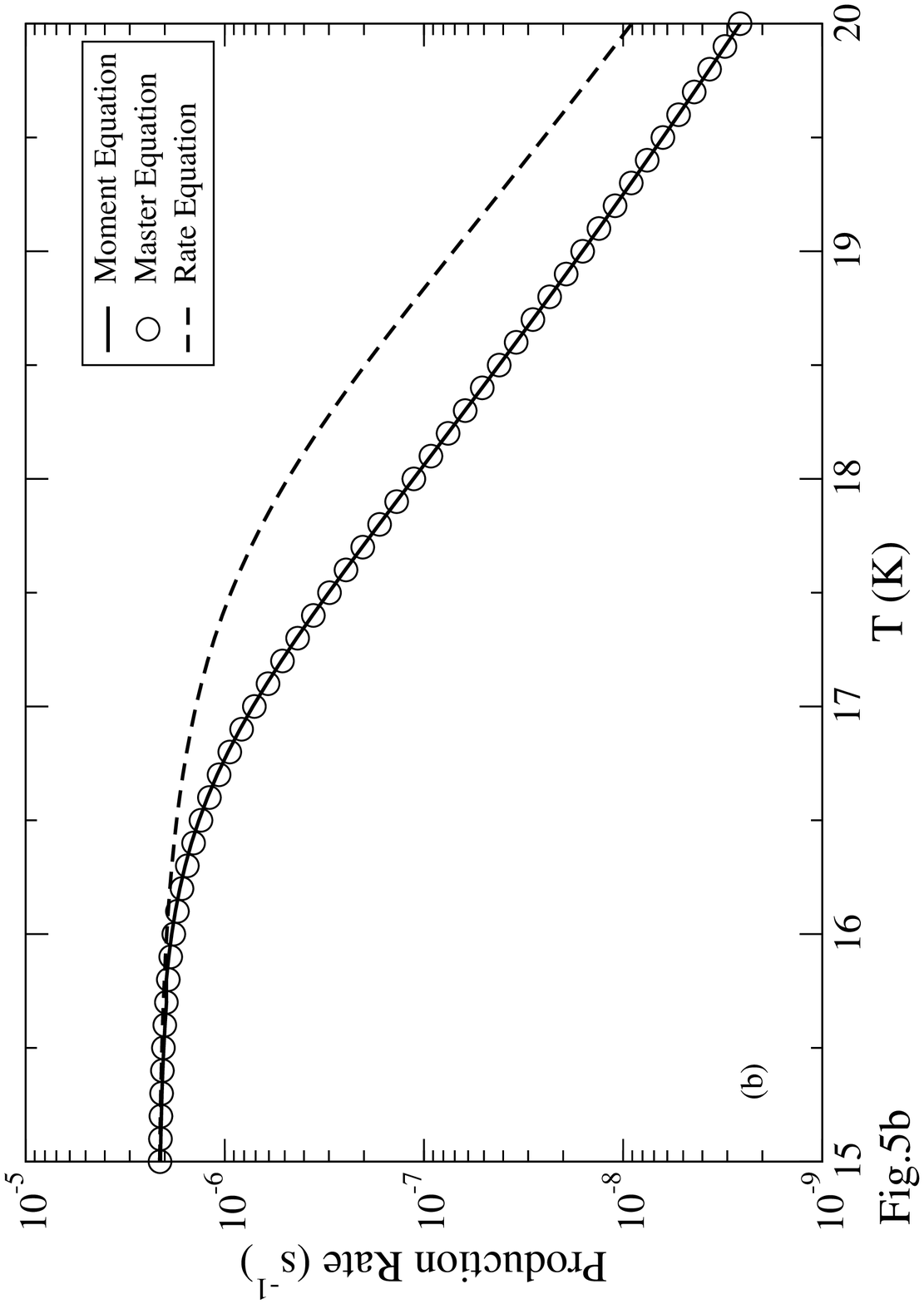}}
\rotr\figeb
\caption{
The average number of H atoms on the grain surface (a), and the production rate
(b) for a grain of radius $r = 10^{-6}$ (cm)
as a function of the grain temperature.
The results of the moment equations 
(solid line)
and the 
master equation
($\bigcirc$) 
coincide perfectly, while the rate equations (dashed line)
significantly overestimate
the production rate for temperatures higher than 16 K.
As the temperature increases above $T = 16$ K, desorption
becomes faster and the production rate decreases. Below 
16 K the production rate saturates and becomes independent
of the temperature within the temperature window of 12 - 16 K.
\label{fig:5}
}
\end{figure}

Observations indicate that the temperatures of dust grains 
in diffuse clouds are typically in the range between 10 - 20 K.
The dependence of the molecular hydrogen formation process 
on the surface temperature of the grains is presented in Fig. 5.
The average number of H atoms on the surface of a very small grain
of radius $r = 10^{-6}$ (cm) is shown in Fig. 5(a). 
The production rate of molecular hydrogen on such grain is shown in
Fig. 5(b).
The results of the moment equations 
(solid line)
and the 
master equation
($\bigcirc$) 
coincide perfectly, while the rate equations (dashed line)
significantly overestimate
the production rate for temperatures higher than 16 K.
As the temperature increases above $T = 16$ K, desorption
becomes faster and the production rate decreases. Below 
16 K the production rate saturates and becomes independent
of the temperature within the temperature window of 12 - 16 K.
Below $T = 12$ K the production rate decreases sharply due to
the lower mobility of H atoms on the surface
\cite{Katz1999}.

\section{Reactions involving two Species}
\label{sec:Complex}

Observations in the past two decades provide evidence that chemical reaction networks
on dust grain surfaces go much beyond the production of molecular 
hydrogen.
In particular, ice mantles that consist of H$_2$O and CO, 
as well as more complex organic molecules,
form on grain surfaces. 

Here we consider a surface reaction network involving hydrogen and oxygen
\citep{Caselli1998}.
In this system surface reactions produce 
H$_2$, O$_2$, OH and H$_2$O molecules.
While the hydrogen molecules typically desorb upon formation
or shortly later, the heavier species tend to remain on the
surface. Some of them, such as OH may then participate in further
reactions.
However, in the model presented below we assume, for simplicity, that
the reaction products do not participate in further chemical reactions
(namely, H$_2$O is not produced).
In practice, we consider these molecules as if they all desorb from
the surface upon formation.

\subsection{Rate equations}

The rate equations that describe the chemical reactions 
on dust grains in our simplified hydrogen-oxygen system are

\begin{eqnarray}
{ {d{\langle N_{\rm H} \rangle}} \over {dt}} & = & 
F_{\rm H} - W_{\rm H} \langle N_{\rm H} \rangle 
- 2 A_{\rm H} {\langle N_{\rm H} \rangle}^{2}  
\nonumber\\ &-&  (A_{\rm H}+A_{\rm O}) \langle N_{\rm H}\rangle 
\langle N_{\rm O} \rangle  \nonumber \\
{ {d{\langle N_{\rm O} \rangle }} \over {dt}} & = & 
F_{\rm O} 
-W_{\rm O} \langle N_{\rm O}\rangle
- 2 A_{\rm O} {\langle N_{\rm O} \rangle}^{2} 
\nonumber\\&-&  (A_{\rm H}+A_{\rm O}) \langle N_{\rm H} \rangle \langle N_{\rm O} \rangle 
\label{eq:NgrainOH} 
\end{eqnarray}

\noindent
where
$N_{\rm O}$ 
is the number of oxygen atoms on the grain
and $A_{\rm O} = a_{\rm O}/S$ is their sweeping rate.
Here we denote the density of the O atoms in the gas phase by
$\rho_O$. 
Their velocity, 
$v_{\rm O}$,
is determined by
Eq. 
(\ref{eq:velocity})
(replacing 
$m_{\rm H}$
by the mass
$m_{\rm O}$
of an oxygen atom)
with the same temperature as that of the hydrogen gas.
The flux of O atoms adsorbed on the grain surface is given by 
$F_{\rm O} = \rho_{\rm O} v_{\rm O} \sigma$ 
(atoms s$^{-1}$).
The desorption rate $W_{\rm O}$ is given by an expression
analogous to
Eq. 
(\ref{eq:P1})
with the same $\nu$
and with 
$E_1$
replaced by the
activation energy for desorption of O atoms. 
The hopping rate 
$a_{\rm O}$
of the O atoms is given by an expression analogous to 
Eq. 
(\ref{eq:Alpha}),
with 
$E_0$
replaced by the
activation energy 
for hopping of O atoms.

As long as the grains are not too small and the fluxes
are not too low, such that 
$N_{\rm H} \gg 1$ 
and
$N_{\rm O} \gg 1$
these rate equations evaluate correctly the formation rates of all
the reaction products. 
However, when the typical numbers of H or O atoms
on the grain go down to order one, these equations become unsuitable
and exhibit significant deviations from the actual production rates.

\subsection{Master equation}

The master equation that describes the hydrogen-oxygen system 
consists of a two dimensional matrix of equations for the time derivatives of
$P_{\rm H\&O}(N_{\rm H},N_{\rm O})$.
It takes the form

\begin{eqnarray}
\dot P_{\rm H\&O}&(&N_{\rm H},N_{\rm O}) = 
F_{\rm H} [P_{\rm H\&O}(N_{\rm H}-1,N_{\rm O}) 
\nonumber\\&-&  P_{\rm H\&O}(N_{\rm H},N_{\rm O})] \nonumber \\ 
&+&F_{\rm O} [P_{\rm H\&O}(N_{\rm H},N_{\rm O}-1) 
\nonumber\\&-& P_{\rm H\&O}(N_{\rm H},N_{\rm O}) ]
\nonumber \\ 
&+& W_{\rm H} [ (N_{\rm H}+1) P_{\rm H\&O}(N_{\rm H}+1,N_{\rm O}) 
\nonumber\\&-&N_{\rm H} P_{\rm H\&O}(N_{\rm H},N_{\rm O})]  \nonumber\\
&+& W_{\rm O} [ (N_{\rm O}+1) P_{\rm H\&O}(N_{\rm H},N_{\rm O}+1) 
\nonumber\\&-& N_{\rm O} P_{\rm H\&O}(N_{\rm H},N_{\rm O})]  \\ 
&+&  A_{\rm H} [ (N_{\rm H}+2)(N_{\rm H}+1) P_{\rm H\&O}(N_{\rm H}+2,N_{\rm O}) 
\nonumber\\&-&  
N_{\rm H}(N_{\rm H}-1)P_{\rm H\&O}(N_{\rm H},N_{\rm O}) ] \nonumber \\
&+&  A_{\rm O} [(N_{\rm O}+2)(N_{\rm O}+1) P_{\rm H\&O}(N_{\rm H},N_{\rm O}+2) 
\nonumber\\&-& 
N_{\rm O}(N_{\rm O}-1) P_{\rm H\&O}(N_{\rm H},N_{\rm O}) ]\nonumber\\
&+&  (A_{\rm H}+A_{\rm O}) 
\nonumber\\&[&(N_{\rm H}+1)(N_{\rm O}+1) P_{\rm H\&O}(N_{\rm H}+1,N_{\rm O}+1) 
\nonumber\\&-& 
N_{\rm H} N_{\rm O} P_{\rm H\&O}(N_{\rm H},N_{\rm O}) ], \nonumber
\label{eq:NmicroOH}
\end{eqnarray}

\noindent
where
$N_{\rm H},N_{\rm O}=0,1,2,\dots$.
In the equations in which 
$N_{\rm H}=0,1$
or 
$N_{\rm O}=0,1$,
some of the terms vanish.
These terms can be easily identified since their
initial or final states include a negative number
of H or O atoms on the grain.

\subsection{Moment equations}

The moment equations for the hydrogen-oxygen system
take the form

\begin{eqnarray}
\frac{d \langle N_{\rm H} \rangle }{dt} &=&
F_{\rm H}+(2A_{\rm H}-W_{\rm H}) \langle N_{\rm H} \rangle 
-2A_{\rm H} \langle N_{\rm H}^2 \rangle  
\nonumber\\&-& (A_{\rm H}
+A_{\rm O}) \langle N_{\rm H}N_{\rm O} \rangle 
\nonumber
\\
\frac{d \langle N_{\rm O} \rangle }{dt} &=&
F_{\rm O}+(2A_{\rm O}-W_{\rm O}) \langle N_{\rm O} \rangle 
-2A_{\rm O} \langle N_{\rm O}^2 \rangle   
\nonumber\\&-& (A_{\rm H}+A_{\rm O}) \langle N_{\rm H}N_{\rm O} \rangle  
\nonumber
\\
\frac{d \langle N_{\rm H}^2 \rangle }{dt} &=&
F_{\rm H}+(2F_{\rm H}+W_{\rm H}-4A_{\rm H}) \langle N_{\rm H} \rangle 
\nonumber\\&+& (-2W_{\rm H} +8A_{\rm H}) \langle N_{\rm H}^2 \rangle   \nonumber \\
&-&4A_{\rm H} \langle N_{\rm H}^3 \rangle 
+(A_{\rm H}+A_{\rm O}) \langle N_{\rm H}N_{\rm O} \rangle 
\nonumber\\&-&2(A_{\rm H}+A_{\rm O}) \langle N_{\rm H}^2N_{\rm O} \rangle 
\nonumber
\\
\frac{d \langle N_{\rm O}^2 \rangle }{dt} &=&
F_{\rm O}+(2F_{\rm O}+W_{\rm O}-4A_{\rm O}) \langle N_{\rm O} \rangle 
\nonumber\\&+&(-2W_{\rm O}+8A_{\rm O}) \langle N_{\rm O}^2 \rangle  \nonumber \\
&-&4A_{\rm O} \langle N_{\rm O}^3 \rangle 
+(A_{\rm H}+A_{\rm O}) \langle N_{\rm H}N_{\rm O} \rangle 
\nonumber\\&-&2(A_{\rm H}+A_{\rm O}) \langle N_{\rm H}N_{\rm O}^2 \rangle 
\nonumber
\\
\frac{d \langle N_{\rm H}N_{\rm O} \rangle }{dt} &=&
F_{\rm H} \langle N_{\rm O} \rangle +F_{\rm O} \langle N_{\rm H} \rangle 
\nonumber\\&+&(-W_{\rm H}-W_{\rm O}+3A_{\rm H}+3A_{\rm O}) \langle N_{\rm H}N_{\rm O} \rangle 
\nonumber\\
&-&(3A_{\rm H}+A_{\rm O}) \langle N_{\rm H}^2N_{\rm O} \rangle 
\nonumber\\&-&(A_{\rm H}+3A_{\rm O}) \langle N_{\rm H}N_{\rm O}^2 \rangle. 
\label{eq:momentOH}
\end{eqnarray}

\noindent
and the production rates of the molecular species 
are given by

\begin{eqnarray}
R_{\rm H_2}^{\rm grain} &=& A_{\rm H} \left(  \langle N_{\rm H}^2 \rangle - \langle N_{\rm H} \rangle \right) \nonumber \\
R_{\rm O_2}^{\rm grain} &=& A_{\rm O} \left(  \langle N_{\rm O}^2 \rangle - \langle N_{\rm O} \rangle \right) \nonumber \\
R_{\rm OH}^{\rm grain} &=& \left(A_{\rm H}+A_{\rm O}\right)  \langle N_{\rm H}N_{\rm O} \rangle.
\end{eqnarray}

\noindent
In the moment equations
the time derivative of  each moment
is expressed as a linear combination of other moments. 
The set can be extended to include higher moments.
However, the number of unknowns (namely moments to be computed) grows
faster than the number of equations. 
In Eqs.
(\ref{eq:momentOH}), 
consisting of five equations, 
there are nine unknowns:  
$\langle N_{\rm H} \rangle $, 
$ \langle N_{\rm O} \rangle $, 
$\langle N_{\rm H}^2 \rangle $,
$ \langle N_{\rm O}^2 \rangle $, 
$ \langle N_{\rm H}^3 \rangle $, 
$ \langle N_{\rm O}^3 \rangle $, 
$ \langle N_{\rm H}N_{\rm O} \rangle $, 
$ \langle N_{\rm H}^2N_{\rm O} \rangle $ 
and 
$ \langle N_{\rm H}N_{\rm O}^2 \rangle $.

Therefore, 
in order to solve the moment equations
one has to use approximations for four of the unknowns listed above.
In principle, these
approximations can be obtained using the following properties.
The number density of hydrogen in the gas phase is by orders of
magnitude higher than that of any other specie, including oxygen.
Therefore, 
$F_{\rm H} \gg F_{\rm O}$.
The activation energies for diffusion and desorption of H atoms
on the grain surface are much lower than those of heavier
atomic species such as oxygen.
Therefore, the diffusion of H atoms on the surface is much faster
and dominates the chemical activity on the surface.
The density of H atoms on the surface is thus primarily controlled
by the processes of H$_2$ formation and H desorption. The further
reduction of H density due to the formation of OH can be considered
as a small correction.
We have examined such approximations and found that they do not
work very well in practice and that the results deviate significantly
from those of the master equation.

\section{Summary}
\label{sec:Summary}

We have introduced a set of moment equations for the analysis of
the formation of molecular hydrogen and other chemical reactions
on dust grain surfaces in the interstellar medium.
These equations are derived from the master equation that describes 
these processes.
Like the master equation, the moment equations
are exact even in the limit of small grains and low
flux where the rate equations fail.
They take into account the discrete nature of the 
populations of atomic and molecular species
that participate in the reactions.
While the master equation is expressed in terms of distributions
such as
$P_{\rm H}(N_{\rm H})$, $N_{\rm H}=0,1,2 \dots$,
the moment equations are expressed in terms of 
the moments of these distributions.

We have shown that the moment equation approach is useful for
simulations of molecular hydrogen formation.
We expect that this approach will make it possible to 
efficiently calculate the total production rate of molecular hydrogen
in diffuse clouds. The results will then be used in order to determine
whether the molecular hydrogen formation process studied here is efficient enough
to account for the observed abundance of H$_2$. 
The input required for such calculations includes the surface
parameters (obtained from laboratory experiments), the grain 
size distribution, the density and temperature of the H gas, 
the grain temperature as well as the dissociation rate of
H$_2$ molecules in the gas phase.
Since the formation rate of molecular hydrogen 
is highly sensitive to the
grain temperature and size distribution,
the accuracy of the results may be limited by the quality of the
observational data. 

For more complex reaction networks involving multiple
species, setting the cutoffs in the moment equations requires
certain approximations that introduce significant errors.
Therefore, for system that consist of more than a single reactive
specie, we currently do not have a useful computational framework
based on the moment equations.
Potential advantages of the moment equations, 
justifying further attempts to further develop this approach, 
include the fact
that they are similar in structure to the rate equations, possibly making it
easier to incorporate them into models of interstellar chemistry.
Moreover, they directly provide the first and second moments,
which quantify the production rates of the molecular species involved.
This is unlike the master equation in which the moments are calculated
from the distribution in a post-processing stage.

\begin{acknowledgements}

We thank E. Herbst, V. Pirronello and G. Vidali for helpful discussions
and correspondence.
This work was supported by the Adler Foundation for Space Research of the
Israel Science Foundation.

\end{acknowledgements}


\clearpage

\end{document}